\input aa.cmm 
%
%
 
%

\def\etal{{\it et al. }}
\def\1{_{\vert}}
\def\refh{\par\noindent\hangindent\parindent\hangafter1}

\def\ueber#1#2{{\setbox0=\hbox{$#1$}%
  \setbox1=\hbox to\wd0{\hss$\scriptscriptstyle #2$\hss}%
  \offinterlineskip
  \vbox{\box1\kern0.4mm\box0}}{}}

\font\tenib=cmmib10
\font\sevenib=cmmib10 at 7pt
\font\fiveib=cmmib10 at 5pt
\newfam\mitbfam
\textfont\mitbfam=\tenib
\scriptfont\mitbfam=\sevenib
\scriptscriptfont\mitbfam=\fiveib

\mathchardef\bfo="0\the\mitbfam21
\mathchardef\om"0\the\bffam0A
\font\bigastfont=cmr8 scaled \magstep 3
\def\bolddot{\hbox{\bigastfont .}}


\MAINTITLE{Averaging inhomogeneous Newtonian cosmologies}

\AUTHOR{Thomas Buchert@1 and J\"urgen Ehlers@2}

\INSTITUTE{
@1 Theoretische Physik, Ludwig--Maximilians--Universit\"at, 
Theresienstr. 37, D--80333 M\"unchen, Germany
@2 Max--Planck--Institut f{\"u}r Gravitationsphysik, Albert--Einstein--Institut,
Schlaatzweg 1, D--14473 Potsdam, Germany}

\DATE{Received ????, accepted ????}
 
\ABSTRACT{
Idealizing matter as a pressureless fluid and representing its motion by a 
peculiar--velocity field superimposed on a homogeneous and isotropic Hubble 
expansion, we apply (Lagrangian) spatial averaging 
on an arbitrary domain $\cal D$ to the (nonlinear) equations
of Newtonian cosmology and derive an exact, general equation for the evolution 
of the (domain dependent) scale factor $a_{\cal D}(t)$. We consider the effect of inhomogeneities on the 
average expansion and discuss under which circumstances the standard description
of the average motion in terms of Friedmann's equation holds. We find that this
effect vanishes for spatially compact models if one averages over the whole 
space. For spatially infinite inhomogeneous models obeying the cosmological 
principle of large--scale isotropy and homogeneity, Friedmann models may provide
an approximation to the average motion on the largest scales, whereas for 
hierarchical (Charlier--type) models the general expansion equation shows how 
inhomogeneities might appreciably affect the expansion at all scales. An 
averaged vorticity evolution law is also given. Since we employ spatial 
averaging, the problem of justifying ensemble averaging does not arise. 
A generalization of the expansion law to general relativity 
is straightforward for the case of irrotational flows and will be discussed.
The effect may have important consequences for a variety of problems in 
large--scale structure modeling as well as for the interpretation of 
observations.}

 
\THESAURUS{
02.07.1; 02.09.1; 03.13.1; 12.03.4; 12.12.1}
\OFFPRINTS{T. Buchert}

\input mssymb
\input psfig
 
\maketitle
 
\titlea{Outline of the problem}
 
Traditionally, cosmological models have been based on the assumption that,
on a sufficiently large scale, the Universe is isotropic and homogeneous.
As long as inhomogeneities are small and described as linear perturbations 
off a Friedmann model which average to zero, this picture is consistent; 
by construction rather than derivation the averaged variables are given by
a Friedmann model. 
However, in the observed part of the Universe the matter inhomogeneities are
not small (e.g., the r.m.s. density fluctuations are 
much larger than unity below the scale of clusters of galaxies), and contemporary theories of structure formation follow the evolution
of inhomogeneities into the nonlinear regime, mostly in the framework of 
Newtonian cosmology, either with analytical approximations or numerical 
N--body simulations. Again, both methods to simulate nonlinear structure 
formation are constructed in such a way that the average flow obeys the 
homogeneous and isotropic Friedmann models. 

Basic, nontrivial questions which are usually not even raised in the 
traditional approach are, whether and how a general anisotropic and 
inhomogeneous solution can be split into an isotropic and homogeneous average
field and deviations thereof and whether, if so, the average variables obey
Friedmann's equation. Because of the nonlinearity of Einstein's as well as
Newton's laws for gravitationally interacting systems, the answers to the last
questions are not obvious, as has been emphasized particularly by Ellis (1984). 

In contemporary models of structure formation the time-- and length--scales
as well as the amplitude of the initial fluctuations are expressed in terms of
quantities of their assumed Friedmann backgrounds such as $a(t)$, $H(t)$,
$\Omega (t)$, etc.$\;$.
Observational data for these parameters are interpreted accordingly.
This procedure which rules
a variety of problems in structure formation theories (like the
question whether nonbaryonic dark matter is needed to explain
present day structure) excludes by assumption all inhomogeneous
models which do not obey Friedmann's equations if averaged on some large scale,
even those which are kinematically isotropic and homogeneous.

In general relativity, the problem of averaging is very 
involved because (i) in a generic spacetime there are
no preferred time--slices one could average over. (It should even be difficult
to recognize a Friedmann model as such if one were given a complicated slice of it.)
(ii) the metric is a dynamical variable entering the field
equations nonlinearly, and it is difficult to average (or instead deform,
see Carfora and Marzuoli 1988, Carfora \etal 1990) it, (iii) a gauge problem arises in relating the
``true'' and the averaged metric. 
Indications that the ``backreaction'' of inhomogeneities 
on the global expansion may have important
consequences for the structure formation process have been put forward by 
Futamase (1989, 1996), Bildhauer (1990) and
Bildhauer \& Futamase (1991a,b) who have studied the ``backreaction'' effect
quantitatively based on perturbative calculations.
In particular, they found that the expansion is
accelerated significantly where inhomogeneities form.
Recently, renormalization group techniques in relation to the averaging problem 
have been advanced by Carfora and Piotrkowska (1995) who investigated in detail the 
connection between (spatial) manifold deformations and spatial averaging
at one instant, using the constraint equations of general relativity.   

In this paper we address the problem of the effect of nonlinear inhomogeneities 
on the average expansion in Newtonian cosmology where also the evolution
equations can be averaged without uncontrolled approximations. 
This provides straightforward insight and some answers 
which are indicative also for the GR--case and uncovers possible limitations of
current cosmological models. Moreover, we will show that for scalar quantities
(like the expansion rate) the averaging procedure proposed in the present
work carries over to the general relativistic case, if we restrict 
the equations to irrotational flows.
 
\titlea{Averages in Newtonian cosmology}

According to Newtonian physics, the motion of a self--gravitat\-ing, 
pressureless fluid (``dust'') is governed by the
{\it Euler--Poisson system} of equations.
Thus, with respect to a non--rotating Eulerian coordinate 
system$^1$\footnote{}{$^1$ notes are given at the end of the paper.} 
the fields of mass density $\varrho(\vec x,t) > 0$, velocity 
$\vec v(\vec x,t)$ and gravitational acceleration
$\vec g(\vec x,t)$ are required to satisfy
$$
\eqalignno {&\partial_t \vec v = - (\vec v \cdot \nabla) \vec v + \vec g  \; ,
&(1a)\cr
&\partial_t \varrho = - \nabla \cdot(\varrho \vec v) \; ,
&(1b)\cr
&\nabla \times \vec g = \vec 0 \; , &(1c) \cr
&\nabla \cdot \vec g = \Lambda - 4 \pi G \varrho \; , &(1d) \cr}
$$
where $\Lambda$ denotes the cosmological constant, here included for
the sake of generality.  
\medskip
It is useful to
introduce the rates of expansion $\theta = \nabla \cdot \vec v$, 
shear $\underline{\sigma}=(\sigma_{ij})$ and 
rotation $\vec\omega = {1\over 2}\nabla
\times \vec v$ 
of the fluid flow via the decomposition\footnote*{a comma denotes partial differentiation
with respect to Eulerian coordinates; we adopt the summation convention.}
$$ 
v_{i,j}=\sigma_{ij}
+{1\over 3}\delta_{ij}\theta + \omega_{ij}\;\;;\;\;\omega_{ij}=
-\eta_{ijk}\omega_k\;\;,\;\;\sigma_{\lbrack i j \rbrack} = 0\;\;,
\eqno(2a)
$$
of the velocity gradient, where $\delta_{ij}$ and $\eta_{ijk}$
denote the (Euclidean) spatial metric and its volume form
(Levi--Civita tensor), respectively. In contrast to $\vec v$, the tensor fields
$\underline{\sigma}$, $\theta$ and $\vec\omega$ are independent of the
(non--rotating) coordinate system; $\vec v$ can be reconstructed from
$\underline{\sigma}$, $\theta$ and $\vec\omega$ up to a spatially constant 
summand (compare Appendix A). 
\medskip
Using the Lagrangian time derivative operator
${d\over dt}\equiv (...)^{\bolddot}=:\partial_t + \vec v \cdot
\nabla$, we may replace the system (1) by the {\it equivalent system} 
(2) consisting of eq. (2a) and the transport equations$^2$
$$
\eqalignno{
\dot{\varrho} &= - \varrho\theta \;\;,&(2b)\cr
\dot{\vec\omega} &= -{2\over 3}{\vec\omega}\theta + 
{\underline{\sigma}}\cdot{\vec\omega}
\;\;,&(2c)\cr
{\dot\theta} &= \Lambda - 4\pi G\varrho - {1\over 3}\theta^2
+2(\omega^2 - \sigma^2 )\;\;,&(2d)\cr}
$$
for $\varrho$, $\vec\omega$ and $\theta$, in which $\sigma$ and $\omega$
denote the magnitudes of shear and rotation, respectively:
$$
\sigma = \vert{\underline{\sigma}}\vert := 
({1\over 2}\sigma_{ij}\sigma_{ij})^{1/2}\;\;,\;\;
\omega = \vert \vec\omega \vert =: 
({1\over 2}\omega_{ij}\omega_{ij})^{1/2}\;\;.
$$
((1) is equivalent to (2) as a system for $\varrho$ and $\vec v$. In fact, if 
(1a) is taken to define $\vec g$ in terms of $\vec v$, then 
$(1c)\Leftrightarrow (2c)$ and $(1d)\Leftrightarrow (2d)$; $(1b)\Leftrightarrow (2b)$ is obvious.)
\medskip
We wish to take spatial averages of the equations (2b,c,d). For this purpose,
let ${\vec f}_t : \vec X \mapsto \vec x$ denote the mapping which takes 
initial positions $\vec X$ of fluid particles at time $t_0$ to their positions
$\vec x$ at time $t$; in other words, let 
$\vec x = {\vec f}_t (\vec X) \equiv\vec f (\vec X,t)$ be the field of trajectories
on which the Lagrangian description of fluid motion is based.
Then, the volume elements at $t$ and $t_0$ are related by 
$d^3 x = J d^3 X$, where $J(\vec X,t): = {D\vec f \over D\vec X}$
is the Jacobian determinant of ${\vec f}_t$. Therefore,
$$
\dot J = J\theta \;\;,\eqno(3)
$$
and the ``comoving'' volume $V(t)=:a_{\cal D}^3 (t)$ of a 
compact portion ${\cal D}(t)$ 
of the fluid changes according to 
$$
\dot V = {d\over dt}\int_{{\cal D}(t)}d^3 x = \int_{{\cal D}(t_0)}d^3 X \;{\dot J}
= \int_{{\cal D}(t)}d^3 x \;\theta\;\;,
$$
which may be written
$$
\langle\theta\rangle_{\cal D} = {\dot V \over V} = 3 {{\dot a}_{\cal D}\over a_{\cal D}}\;\;.
\eqno(4)
$$
Here and in the sequel, $\langle A\rangle_{\cal D}=
{1\over V}\int_{\cal D} d^3 x \, A$ 
denotes the spatial average of a (spatial)
tensor field $\cal A$ on the domain ${\cal D}(t)$ occupied by the amount of fluid 
considered, and $a_{\cal D}(t)$ is the scale factor of that domain.
Suppressing temporarily the index ${\cal D}$ for notational simplicity, 
we note the useful ``commutation rule''
$$
\langle {\cal A} \rangle^{\bolddot} - \langle {\dot {\cal A}} \rangle = \langle {\cal A} 
\theta \rangle - \langle {\cal A} \rangle \langle \theta \rangle \;.\eqno(5)
$$
(Proof:

\noindent
$\langle {\cal A} \rangle^{\bolddot} = {d\over dt}(V^{-1} \int d^3 x \;{\cal A} ) = 
-{{\dot V} \over V}\langle {\cal A} \rangle
+V^{-1}\int d^3 X \; ({\dot {\cal A}} J + {\cal A} {\dot J})$; using (3) and (4) gives (5).)
\medskip
Applied in turn to $\varrho$, $\vec\omega$ and $\theta$, eq. (5) combined with
(2b,c,d), respectively, leads to
$$
\langle\varrho\rangle^{\bolddot} = - \langle\varrho\rangle\langle\theta\rangle
\;\;,\eqno(6)
$$
$$
\langle\vec\omega\rangle^{\bolddot}=\langle\vec\omega \cdot \nabla \vec v 
\rangle - \langle\vec\omega \rangle\langle\theta\rangle\;\;,
\eqno(7)
$$
$$
\langle\theta\rangle^{\bolddot} = \Lambda - 4\pi G \langle\varrho\rangle +{2\over 3}(\langle\theta^2 \rangle) - \langle\theta\rangle^2 
+2 (\langle\omega^2 \rangle-\langle\sigma^2 \rangle)\;\;,\eqno(8)
$$
where we have also used eq. (2a). 
\medskip
In terms of the scale factor $a$, eq. (6) says that $a^3 \langle\varrho\rangle =:M$ is the
constant total mass of the fluid portion considered.
Eq. (8) may be rewritten as
$$
3{\ddot a \over a} + 4\pi G {M\over a^3 } -
\Lambda = {2\over 3}\left(\langle\theta^2 \rangle - \langle\theta\rangle^2 \right) +
2\langle\omega^2 - \sigma^2 \rangle\;\;.\eqno(9a)
$$
The averaged Raychaudhuri equation (9a) can also be written 
as a standard Friedmann equation 
for the ``effective mass density'' $\varrho_{\rm eff}$,
$$
4\pi G \varrho_{\rm eff}:= 4\pi G \langle\varrho \rangle  
- {2\over 3}\langle (\theta  - \langle\theta \rangle )^2 \rangle +
2\langle\sigma^2 - \omega^2 \rangle\;\;.\eqno(9b) 
$$
Eq. (9) shows that inhomogeneities have an accelerating effect
on the averaged volume expansion rate $\langle\theta\rangle$, 
if the shear term dominates the
vorticity and contraction terms in that equation.
Eqs. (9) can also be used to discuss anisotropies: If the domain $\cal D$ is
chosen as a cone with the observer at the vertex, (9a) shows that the average
expansion in two such cones will be different if the averages involving 
$\theta$, $\omega$ and $\sigma$ differ.
\smallskip 
\noindent 
From eq. (7) we conclude that 
$$
\langle{\vec\omega}\rangle = \vec 0 \;\;\Rightarrow\;\;
\langle{\vec\omega}\cdot\nabla\vec v \rangle
=\vec 0
\;\;.\eqno(10a)
$$
The converse only holds asymptotically for positive average expansion: 

\noindent
if $\langle{\vec\omega}\cdot\nabla\vec v \rangle = \vec 0$, then
$$
\langle{\vec\omega}\rangle=\langle{\vec\omega}(t_0)\rangle\,
e^{-\int_{{\cal D}(t)}dt\,\langle\theta \rangle \;(t-t_0)} \;\;;\eqno(10b)
$$
for a universe which collapses on average, $\langle{\vec\omega}\rangle$ 
blows up (compare Olson \& Sachs 1973)$^3$.

\bigskip

The equations (6)--(9) hold for any part of any pressureless fluid. 

\bigskip
We want to apply the foregoing results
to a large, typical part of a cosmological model. For this purpose we split
the velocity field into a {\it Hubble flow} and a {\it peculiar--velocity}
$\vec u$. Recall that a Hubble flow 
can be characterized as having vanishing rates of shear and
rotation and a spatially constant rate of expansion $3H(t)$ (cf. eq. (2a)). 
Therefore,
$$
v_{i,j} = H(t)
\delta_{ij} + u_{i,j}\;\;,\eqno(11a)
$$
hence
$$
\theta = 3H + \nabla \cdot \vec u \;\;.\eqno(11b)
$$
We may choose$^4$
$$
H:={{\dot a}_{\cal D} \over a_{\cal D}}\;\;,\eqno(12a)
$$
or, equivalently,
$$
\langle \nabla \cdot \vec u \rangle_{\cal D} = 0 \;\;.\eqno(12b)
$$
$H(t)$ and $\vec u$ then depend, of course, 
on the choice of the averaging region $\cal D$.

From eqs. (11a,b) and (2a) we get by a little computation 
an expression for the quadratic principal 
invariant of the velocity gradient $(v_{i,j})$:
$$
2 {\bf II}(v_{i,j}) \equiv
{2\over 3}\theta^2 + 2(\omega^2 - \sigma^2 ) 
$$
$$
= 6H^2 + 4H\,\nabla \cdot \vec u +
\nabla\cdot\left(\vec u \,\nabla\cdot \vec u - \vec u \cdot
\nabla \vec u \right)\;\;.\eqno(13)
$$
We use it and eq. (12a) to rewrite the {\it averaged Raychaudhuri equation} (9a) 
in the form
$$
3{{\ddot a}_{\cal D} \over a_{\cal D}} + 4\pi G {M_{\cal D}\over a_{\cal D}^3} -
\Lambda \;=\; 
a^{-3}_{\cal D}\int_{\partial{\cal D}(t)}\vec{dS}\cdot\left(\vec u 
\,\nabla\cdot \vec u - \vec u \cdot
\nabla \vec u \right),
$$
$$
\eqno(14)
$$
where we have reinserted the index $\cal D$ 
to exhibit the dependence on the comoving domain
chosen, and we have applied Gau{\ss}'s theorem to transform 
the volume integral
in the average to a surface integral over the boundary $\partial{\cal D}$ of
the domain. 
\medskip
We note that the {\it averaged Helmholtz vorticity equation} (7) can similarly be brought
into a more transparent form. Since the Hubble flow is irrotational, we have 
$\vec\omega = {1\over 2}
\nabla \times \vec u$; then, using Green's formula,
$$
\langle\vec\omega \rangle_{\cal D} = {1\over 2}a^{-3}_{\cal D}\int_{\partial{\cal D}(t)}
\vec{dS}\times\vec u\;\;,
\eqno(15)
$$
$$
(\langle\vec\omega \rangle_{\cal D})^{\bolddot} + 2H\langle\vec\omega \rangle_{\cal D} = 
\langle\vec\omega\cdot\nabla \vec u \rangle_{\cal D} -
\langle\vec\omega \rangle_{\cal D}a^{-3}_{\cal D}
\int_{\partial{\cal D}(t)}\vec{dS}\cdot\vec u \;\;.
$$
$$
\eqno(16)
$$
(Olson \& Sachs (1973) have derived an evolution equation for the ensemble
average of ${\omega}^2$ assuming homogeneous--isotropic turbulence.)
\bigskip
We emphasize that, given a solution $(\varrho,\vec v )$ of the basic equations 
(1),
the terms in eqs. (12)--(16) depend on the choice of an ``initial'' domain ${\cal D}(t_0)$
and a choice of a Hubble function $H(t)$.
\medskip
So far, all equations refer to a single non--rotating, rectangular Eulerian coordinate system
$x^a$. Such a description holds globally, if space is
assumed to be the standard Euclidean ${\Bbb R}^3$.
However, we wish to consider also spatially compact Newtonian models as defined and 
analyzed in (Brauer \etal 1994). As shown there, space can then be considered without loss
of generality as a torus ${\Bbb T}^3$, which 
cannot be covered by a single coordinate system. It is important 
(especially for Subsection 3.1) that never\-theless the
eqs. (14)--(16) remain valid for such models as intrinsic, coordinate--free relations for any
compact, orientable domain ${\cal D}(t_0)$ and its evolution ${\cal D}(t)$,
provided the models admit a Hubble flow.
The reason is that, although the inhomogeneous fluid motion (locally given by 
$\vec v$)
as well as the Hubble flow cannot be described globally 
by smooth vector fields on the torus,
the peculiar--velocity $\vec u$, being the {\it relative} velocity of those 
two motions, {\it is} a smooth, global $3-$vector field on ${\Bbb T}^3$
(see Appendix A for a detailed discussion of the kinematics on toroidal models).

\titlea{Interpretation for different cosmologies}

Let us now assume that the Hubble flow 
represents the mean motion on some domain $\cal D$ and, correspondingly, 
that the peculiar--velocity field $\vec u$ represents the inhomogeneous 
flow with $\langle u_{i,j}\rangle_{\cal D} = 0$ which, by eq. (11a), 
is equivalent to eq. (4) and $\langle \sigma_{i,j}\rangle_{\cal D} = 0$,
$\langle \omega_{i,j}\rangle_{\cal D} = 0$.
Eq. (14) then shows how this field $\vec u$ determines the deviation
of the expansion, on the scale $a_{\cal D}$, from Friedmann's law.
The ``perturbing terms'' are quadratic in $\vec u$ and are 
in general nonzero (even for isotropic expansion); a similar 
remark applies to eq. (16).

\medskip\noindent
We now consider three theoretically possible cosmologies.

\titleb{Spatially compact cosmologies}

These models with a {\it closed} (i.e., compact without boundary) 3--space 
admit either none or exactly one (global) Hubble flow, for
the existence of such a flow means that the toroidal space expands 
self--similarly without rotation (see Appendix A). In this latter case we may choose
the compact domain ${\cal D}(t)$ to be the whole space. Then, ${\cal D}(t)$
has no boundary, and we obtain from eqs. (11a) and (14)--(16):
$$
3{\ddot a\over a} + 
4\pi G {M_{tot}\over a^3} -\Lambda = 0\;\;,\;\;
$$
$$
\langle\sigma_{ij}\rangle = 0 \;\;;\;\;
\langle\vec\omega \rangle = \vec 0 \;\;,\;\;\langle\vec\omega \cdot \nabla \vec u \rangle = 
\vec 0 \;\;,\eqno(17a)
$$
and, from (9a) and (14),
$$
{1\over 3}\langle (\theta  - \langle\theta \rangle )^2 \rangle +
\langle \omega^2 \rangle\;=\;\langle\sigma^2 \rangle\;\;.\eqno(17b) 
$$
(Here we have used that $\vec u$ is globally well--defined.)
\medskip
For inhomogeneous models based on such Friedmann backgrounds
Brauer \etal (loc.cit.) have established some exact qualitative results,
and in (Ehlers \& Buchert 1996) we develop a perturbation scheme the terms of which
are well--defined  to arbitrary order and are uniquely determined by initial data;
that scheme is also useful for numerical work.
\medskip
Toroidal models with a background Hubble flow are, in effect, 
used in all current structure formation simulations,
since they employ periodic boundary conditions for the density and peculiar--velocity
fields on some large scale.
As we have just shown, such (Newtonian) models can always be regarded as finite perturbations
of toroidal
Friedmann models. The perturbations, whether small or large, cannot influence the
overall expansion.
 
\titleb{Cosmologies based on the principle of `large--scale homogeneity 
and isotropy'}

If we take space to be infinite Euclidean ${\Bbb R}^3$, we can assume the boundary term
in (14) to be small compared to ${{\ddot a}_{\cal D} \over a_{\cal D}}$. 
This may be the case if the inhomogeneities are substantially smaller
than the size $a_{\cal D}$ of the averaging domain and if 
their peculiar--velocities are small on the averaging scale
($\vert\vec u \vert << a_{\cal D} H$); 
it corresponds to the cosmological principle of
`statistical homogeneity and isotropy'. Then, the average motion may be 
{\it approximately} given by a Friedmann model on a scale which is considerably larger
than the largest existing inhomogeneities. This is the generally held view.

\titleb{Hierarchical (Charlier--type) cosmologies}

Observational evidence for structures on scales of hundreds of Megaparsecs 
indicates that the ``closure scale'' -- the circumference of the Universe if it
is closed -- has certainly not been reached   
by observed samples of the Universe.
Power spectrum estimates of the density contrast show
a negative slope close to the limit of the deepest available surveys.
However, the COBE microwave background measurement
suggests that the spectrum bends over to positive slopes on larger
scales, supporting the assumption of large--scale homogeneity.
Thus, if the COBE measurement is indeed a detection of {\it primordial}
density fluctuations,
then (up to the cosmic variance of this measurement) we may use 
Friedmann models as approximate models for the average flow on the largest
scales.

The expansion law (14), however, indicates that Friedmann's law may be appreciably
modified by peculiar--velocity terms; for large--scale homogeneity and isotropy
the sum of these terms will approach a limiting value which may or may not be
negligible, in contrast to the spatially compact case. 
Only in the case where this limiting value is negligible, 
global values of, e.g., the Hubble constant
and the density parameter are related to the Friedmann parameters on that 
``asymptotic'' scale. If such a scale doesn't exist, 
the average flow will evolve into an anisotropic and (after development of multi--streaming)
rotational flow, in spite of isotropic and irrotational initial data.
The extreme opposite, a (globally) 
{\it hierarchical cosmology}, where the spectrum of fluctuations
continues to rise on large scales, could {\it only} be treated by 
including the source terms. 

\titleb{Supplementary remarks}

Eqs. (9a) and (14) imply that the expression 
$$
{\cal Q}: = 
{2\over 3}\langle (\theta  - \langle\theta \rangle )^2 \rangle +
2\langle \omega^2 \rangle - 2\langle\sigma^2 \rangle\eqno(18) 
$$
is a divergence. From this we conclude:
\smallskip\noindent
i) If the shear vanishes in a toroidal model which admits a background
Hubble flow, the model is homogeneous and isotropic; in other words:
\smallskip\noindent
ii) In order to contain any inhomogeneities at all, compact models 
admitting a background Hubble flow must have nonvanishing shear. 
This statement applies to the shear scalar, the shear tensor 
as well as to the 
average $\langle\sigma^2 \rangle$.
\smallskip\noindent 
iii) If, on {\it all} sufficiently large scales -- e.g., because of the
structure formation process -- the expression $\cal Q$ in nonzero, then the
model either is not compact or does not admit a background Hubble flow.
\smallskip\noindent
iv) Even if the expression $\cal Q$ is zero on some large scale $A$, then 
still $\langle \sigma^2 \rangle_A \ne 0$ and, in general, $\theta \ne \langle \theta 
\rangle_A$, provided inhomogeneities are present. 
\medskip\noindent
These four statements are equivalent; they only emphasize different aspects of
i). To prove i), note that if $\sigma = 0$ and the model is compact, the
integral over the nonnegative function 
${2\over 3}\langle (\theta  - \langle\theta \rangle )^2 \rangle +
2\langle \omega^2 \rangle$ vanishes; hence $\theta = \langle \theta \rangle$
and $\langle {\vec\omega} \rangle = \vec 0$, therefore, $\vec v$ itself represents
a Hubble flow, q.e.d.

Note that Remarks ii) and iv) concerning the shear follow, 
because the average over $\sigma^2$ is performed over non--negative
terms and can only vanish, if the shear scalar vanishes pointwise and, hence,
if each component of the shear tensor itself is zero. Turning this argument
around, nonvanishing components of the shear tensor result in a nonzero
average of $\sigma^2$ on the scale $A$.
Since this average does not vanish, 
the average fluctuation of the expansion does neither, unless the average 
vorticity exactly compensates the average shear (see also Yodzis 1974).
In general we neither expect the summands in (18) to vanish individually,
nor to decrease by going to larger volumes$^5$. This indicates that 
the ``conspiracy assumption'' ${\cal Q} = 0$ on some large scale 
(if the Universe is not genuinely compact) must be considered a strong 
restriction of generality. 
The existence of a Hubble flow (as a reference flow) does not 
imply that the {\it average flow} is a Hubble flow.

\bigskip
As we have indicated in the introduction, the averaging problem in 
general relativity is very involved. However, for scalar quantities like
the expansion rate $\theta$ we may derive
an expansion law in full analogy to the Newtonian case. For this purpose
we may foliate the spacetime into a family of space--like hypersurfaces 
with spatial metric $g_{ij}$
which are flow--orthogonal (this is only possible, if the flow is irrotational).
Introducing Gaussian (normal) coordinates we can define spatial averages 
(e.g. the average of $\theta$) in a spatial domain $\cal D$ with volume 
$V=\int_{\cal D} d^3 X \sqrt{g}$, $g:=\det(g_{ij})$, as follows:
$$
\langle \theta \rangle_{\cal D} : = 
{1\over V}\int_{\cal D} \theta \sqrt{g} d^3 X \;\;\;.\eqno(19)  
$$
This definition agrees with equation (4), if $J$ is replaced by $\sqrt{g}$.
(Indeed, if the metric is written as the quadratic form
$g_{ij}=\eta^a_{\;i} \eta^a_{\;j}$ involving the one--forms 
${\bf\eta}^a = \eta^a_{\;i}dX^i$, 
then $\sqrt{g} =\det(\eta^a_{\;i})= :J$. $J$ is identical to the Jacobian 
determinant used in the present work, if $\eta^a_{\; i} \equiv \nabla_0 {\vec f}$.)
Since Raychaudhuri's equation is the same in GR we conclude that the 
expansion law (9) is also valid in general relativity.

Yodzis (1974) has shown (Theorem 2) that, for  
general--relativistic, irrotational
dust models for which the hypersurfaces orthogonal to the
dust world lines are closed and orientable, equation (9) 
holds if the averages are performed over the whole space. Our statement is more
general in the sense that equation (9) holds in general relativity
for the same assumptions, but for arbitrarily chosen 
spatial domains. 

An important difference
to the Newtonian treatment, besides spatial curvature, arises due to the fact
that it may not be in general possible to represent the term (18) 
as a divergence in GR.
We stress that this would imply a strong challenge for the standard cosmologies,
since we can no longer argue, except for non--generic situations, that there
exist cases in which the average obeys Friedmann's law. Even more, we don't 
expect the previously discussed arguments (after eq. (18)) to hold, 
since the valid theory on the large scales under consideration 
is general relativity.   

\bigskip

\acknow{
We would like to thank Bernard Jones and Herbert 
Wagner for valuable discussions and suggestions.
TB is supported by 
the ``Sonderforschungsbereich 375 f\"ur
Astro--Teilchenphysik der Deutschen Forschungsgemeinschaft''
and acknowledges hospitality offered by the Albert--Einstein--Institut during working visits.}

\titlea{Notes}

\refh $1)\,$ For spatially unbounded mass distributions there are no inertial
coordinate systems related by Galilean transformations. Instead, the 
preferred coordinate systems ($t,x^{a}$) are related by the more general 
transformations
$x^{a'} = R^{a'}_{\;b} x^b + d^{a'} (t)$, 
where $R^{a'}_{\;b}$ denotes a constant rotation matrix and $d^{a'} (t)$
represents a translation depending arbitrarily on time $t$. The basic 
equations (1) are covariant with respect to these transformations, provided
$\varrho$, $\vec v$ and $\vec g$ are transformed in the obvious way.
These coordinate systems are called (dynamically) non--rotating, since with 
respect to them, no Coriolis forces occur (Heckmann and Sch\"ucking
1955,56).

\refh $2)\;$ We call eqs. (2b,c,d) transport equations since they describe how 
$\varrho$, $\vec\omega$, $\theta$ change along a fluid trajectory. In 
Lagrangian coordinates $(t,\vec X )$ eqs. (2) are ordinary differential
equations with $\vec X$ as parameters.

\refh $3)\;$ The assumption of homogeneous--isotropic turbulence
employed by Olson \& Sachs (1973) imposes from the start a strong restriction
which excludes the relevant terms discussed here. Also, they use ensemble 
averaging instead of spatial averaging. Of course, statistical
statements for spatial averages can be investigated by averaging over
statistical ensembles of spatial domains ${\cal D}$ centered at random 
points in space.

\refh $4)\;\,$ Given $v_i$ and $H(t)$, (11a) defines $\vec u$ up to a 
spatially constant term. One can fix $\vec u$ uniquely by requiring either
$\langle \vec u \rangle = \vec 0$ or $\langle \varrho \vec u \rangle = \vec 0$.
A weaker assumption than eq. (12b) 
is to require $\langle\nabla \cdot \vec u \rangle = 0$ only on the largest scales. In this
case $H(t)$ could be defined in terms of a Friedmann solution, whereas
$H_{\cal D} : = {{\dot a}_{\cal D} \over a_{\cal D}}$ could be interpreted as
the value of Hubble's constant as measured on smaller scales (after averaging
over statistical ensembles of spatial domains). The inhomogeneous term on the
r.h.s. of eq. (14) would then contain additionally a surface integral of the
flux of $\vec u$ through the boundary $\partial{\cal D}$. Since
$a_{\cal D}$ is defined via the volume, it could be interpreted also as an
``effective'' ({\cal D}--dependent) scale factor of a possibly 
anisotropically expanding domain (compare also Appendix B and, for 
further discussion, Buchert (1996)).

\refh $5)\;$ Imagine we divide a large portion of the Universe $\cal D$ 
with volume $V(t)$ into $N$ subdomains ${\cal D}_i$ with volumes 
$V_i (t)$.
Then, e.g., $\langle\sigma^2 \rangle_{\cal D} = 
{1\over N}\sum_i \langle\sigma^2 \rangle_{{\cal D}_i}$. 
If the subdomains are `typical', then we conclude
$\langle\sigma^2 \rangle_{\cal D} = \langle\sigma^2 \rangle_{{\cal D}_i}$, i.e.,
the average value of $\sigma^2$ attained in the subdomains is ``frozen''
and cannot become smaller by averaging over larger domains.
This statement also applies to any positive semi--definite quantity
such as $\omega^2$ and $(\theta - \langle\theta\rangle)^2$.     

\bigskip\medskip
 
\begref{References}

\ref
Bildhauer S., 1990, Prog. Theor. Phys. 84, 444
\ref
Bildhauer S., Futamase T., 1991a, MNRAS 249, 126
\ref
Bildhauer S., Futamase T., 1991b, GRG 23, 1251
\ref
Brauer U., Rendall A., Reula O., 1994, Class. Quant. Grav. 11, 2283
\ref
Buchert T., 1996, in ``Mapping, Measuring and Modelling the Universe'',
Val\`encia 1995, P. Coles \etal (eds), ASP conference series, pp. 349--356 
\ref
Carfora M., Marzuoli A., 1988, Class. Quant. Grav. 5, 659
\ref
Carfora M., Isenberg J., Jackson M., 1990, J. Diff. Geom. 31, 249
\ref
Carfora M., Piotrkowska K., 1995, Phys. Rev. D 52, 4393
\ref
Ehlers J., Buchert T., 1996, GRG, submitted
\ref
Ellis G.F.R., 1984, in Proc. 10th international conference on 
General Relativity and Gravitation, B. Bertotti \etal (eds), Reidel Dordrecht
\ref
Futamase T., 1989, MNRAS 237, 187
\ref
Futamase T., 1996, Phys. Rev. D 53, 681
\ref
Heckmann O., Sch\"ucking E., 1955, Zeitschrift f\"ur Astrophysik 38, 95 
(in German)
\ref
Heckmann O., Sch\"ucking E., 1956, Zeitschrift f\"ur Astrophysik 40, 81 
(in German)
\ref
Olson D.W., Sachs R.K., 1973, ApJ 185, 91
\ref
Yodzis P., 1974, Proc. Roy. Irish Acad. 74A, 61

\endref

\vfill\eject

\appendix{$\;$A:$\;$Kinematics on toroidal spacetimes}

\noindent
Consider any ``Newtonian'' toroidal spacetime ${\Bbb S}={\Bbb T}^3 \times 
{\Bbb R}\,$.
The ``global motion'' of the space ${\Bbb T}^3$ of $\Bbb S$ is given by a
time--dependent triad {${\vec e}_i (t)$} of spatial vectors
generating those translations $n^i {\vec e}_i (t)$ which map each point of
the torus ${\Bbb T}^3$ onto itself at time $t$, $n^i$ being arbitrary integers.
The rates of change ${\dot{\vec e}}_i$, measured with respect to a
non--rotating orthonormal basis, uniquely determine a matrix $H_{ij}(t)$
via  ${\dot{\vec e}}_i = H_{ij}{\vec e}_j$.

With respect to any particle as origin, any point of ${\Bbb T}^3$ has 
infinitely many position vectors $\vec x + n^i {\vec e}_i (t)$ at time $t$.
Hence, any kinematically possible motion covering the universe $\Bbb S$ can be
represented by a velocity field $\vec v (\vec x,t)$ on the covering spacetime
${\Bbb R}^3 \times {\Bbb R}$ of ${\Bbb S}$ which obeys the {\it torus condition}
$$
\vec v (\vec x + n^i {\vec e}_i (t),t) = \vec v (\vec x,t) + 
n^i {\dot{\vec e}}_i (t) \;\;.\eqno(T)
$$
It relates the velocities associated with the position vectors of any
one particle. The difference of two such velocity fields, {\it but not these
fields themselves}, define global spatial vector fields on the torus, as (T)
shows. This is illustrated in the figure below:

\vskip 0.5 true cm
\centerline{\psfig{figure=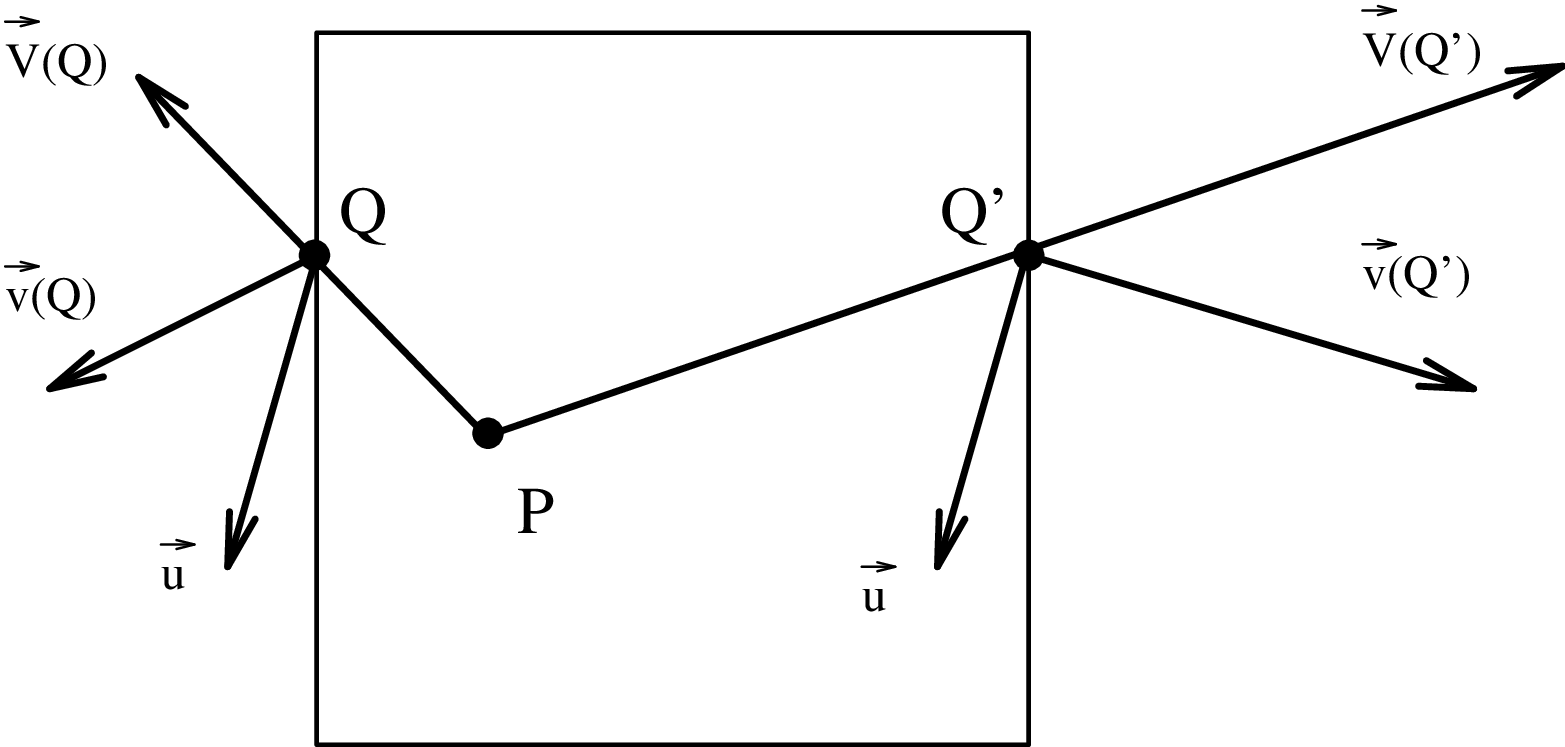,height=3.5cm,width=7.3cm}}
\vskip 0.2 true cm

If $\vec v$ obeys (T), then $v_{i,j}(\vec x + n^i {\vec e}_i (t),t) = 
v_{i,j} (\vec x,t)$, hence $\sigma_{ij}$, $\omega_{ij}$ and $\theta$ 
{\it are} tensor fields on ${\Bbb T}^3$. 

A velocity field $\vec V$ 
obeys the cosmological homogeneity principle iff, for all $\vec y$ and $\vec x$,
$$
\vec V (\vec y,t) - \vec V (\vec x,t) = A(t) (\vec y - \vec x) \eqno(C)
$$ 
(Heckmann \& Sch\"ucking 1955, 1956). This property is compatible with (T) iff
$A = H$, as follows by putting $\vec y = \vec x + n^i {\vec e}_i (t)$ in 
equation (C) and using (T). This shows: Except for a spatially constant
additive contribution, there exists exactly one homogeneous motion on a 
toroidal spacetime; a representative of its velocity field is determined by
the matrix $H_{ij}$ defined above, as $\vec V (\vec x,t) = H(t)\vec x$. Its
shear $\Sigma_{ij}$, rotation $\Omega_{ij}$ and expansion $\Theta$ are 
spatially constant and given by the decomposition 
$$
H_{ij} = \Omega_{ij} + \Sigma_{ij} + {1\over 3}\Theta \delta_{ij}\;\;.
$$

\noindent
If $\vec v$ is any inhomogeneous flow on $\Bbb S$ and $\vec V + A(t)$ the
general homogeneous one, then a unique peculiar--velocity field $\vec u$
is defined by $\vec u : = \vec v - \vec V$, with 
$\langle \vec u \rangle_{{\Bbb T}^3} = \vec 0$. 
Since $\vec u$ is a global vector field on ${\Bbb T}^3$, $\langle u_{i,j} 
\rangle_{{\Bbb T}^3} = 0$, hence
$$ 
\langle\omega_{ij}\rangle_{{\Bbb T}^3} = \Omega_{ij} \;\;,\;\;
\langle\sigma_{ij}\rangle_{{\Bbb T}^3} = \Sigma_{ij} \;\;,\;\;
\langle\theta\rangle_{{\Bbb T}^3} = \Theta \;\;.
$$
The homogeneous motion of a toroidal model is a Hubble flow if and only if
$\Sigma_{ij} = 0$ and $\Omega_{ij} = 0$ or, equivalently, 
$H_{ij} = {1\over 3}\Theta\delta_{ij}$; the last condition is in turn 
equivalent to ${\vec e}_i (t) = {a(t)\over a(t_0)}{\vec e}_i (t_0)$, if we
put $\Theta = 3{\dot a \over a}$. 

\noindent
Thus, a Hubble flow exists and is then 
uniquely determined on a toroidal model if and only if the torus expands
self--similarly without rotation.

\appendix{$\;$B:$\;$General expansion law with global shear and vorticity}

\noindent
We start from Raychaudhuri's equation, averaged 
on a simply connected domain (eq. (9a)), 
$$
3{{\ddot a}_{\cal D} \over a_{\cal D}} + 4\pi G {M\over a_{\cal D}^3 } -
\Lambda = {2\over 3}\left(\langle\theta^2 \rangle_{\cal D} - 
\langle\theta\rangle_{\cal D}^2 \right) +
2\langle\omega^2 - \sigma^2 \rangle_{\cal D} \;\;.
$$ 
By introducing on $\cal D$ arbitrary fields of expansion $\Theta = \Theta (t)$,
shear $\Sigma_{ij} = \Sigma_{ij} (t)$ and vorticity 
$\Omega_{ij} = \Omega_{ij} (t)$
(which are not necessarily average values, but merely 
define a time--dependent standard of reference, e.g., a homogeneous
solution of the basic system of equations),
we may define a linear ``background velocity field'' $\vec V$ 
by $V_i := H_{ij}x_j$ with 
$$
V_{i,j} =\Sigma_{ij}+{1\over 3}\Theta \delta_{ij} + 
\Omega_{ij} = : H_{ij}\;\;.\eqno(B.1)
$$
Also, the inhomogeneous deviations from these reference values are introduced,
$$
\theta = \Theta + \hat{\theta} \;\;;\;\;\sigma_{ij} = \Sigma_{ij} + \hat{\sigma}_{ij}
\;\;;\;\;\omega_{ij} = \Omega_{ij} + \hat{\omega}_{ij}\;\;.\eqno(B.2a,b,c) 
$$ 
Equation (9a) may then be cast into the form
$$
3{{\ddot a}_{\cal D} \over a_{\cal D}} + 4\pi G {M\over a_{\cal D}^3 } -
\Lambda = 
$$
$$
2\left(\Omega^2 - \Sigma^2 \right) 
+ 2\left(\Omega_{ij}\langle\hat{\omega}_{ij}\rangle_{\cal D}
- \Sigma_{ij}\langle\hat{\sigma}_{ij}\rangle_{\cal D}\right)
$$
$$
+ {2\over 3}(\langle{\hat\theta}^2 \rangle_{\cal D} - 
\langle\hat{\theta}\rangle_{\cal D}^2 ) +
2\langle\hat{\omega}^2 - \hat{\sigma}^2 \rangle_{\cal D}\;.
$$
$$
\eqno(B.3)
$$
Using (13) to express deviations from homogeneity in terms of 
the peculiar--velocity gradient $(u_{i,j})$ with 
$H^2 = H_{ij}H_{ij} = 2(\Sigma^2 + \Omega^2 ) +{1\over 3}\Theta^2$, 
we arrive at the (most general) expansion law which holds for 
general inhomogeneities relative to a non--isotropic and 
rotational ``background velocity field'':
$$
3{{\ddot a}_{\cal D} \over a_{\cal D}} + 4\pi G {M\over a_{\cal D}^3 } -
\Lambda = 
$$
$$
2\left(\Omega^2 - \Sigma^2 \right) + 
2\left(\Omega_{ij}\langle\hat{\omega}_{ij}\rangle_{\cal D}
- \Sigma_{ij}\langle\hat{\sigma}_{ij}\rangle_{\cal D}\right)
$$
$$
+ \langle\nabla\cdot \lbrack \vec u (\nabla \cdot \vec u ) - 
(\vec u \cdot \nabla) \vec u \rbrack \rangle_{\cal D} 
- {2\over 3} \langle\nabla \cdot \vec u \rangle_{\cal D}^2 \;\;. 
\eqno(B.4)
$$

\bigskip\noindent
Instead of considering inhomogeneities relative to 
a ({\cal D}--inde\-pen\-dent) arbitrary reference velocity
field $\vec V$, we may alternatively define the inhomogeneous 
fields in (B.2a,b,c) 
as deviations from the average flow within the domain $\cal D$
(which we identify with the global background
velocity field). This corresponds to the point of view of an observer who
maps a finite region of space and assumes that the average values of that 
region are `typical' for the Universe.
In this case all global variables $\Theta (t)$, 
$\Sigma_{ij} (t)$ and 
$\Omega_{ij} (t)$ are averages and depend on content, shape and position 
of the spatial domain $\cal D$:
$$
\Theta  = \langle\theta\rangle_{\cal D} \;\;;\;\;\Sigma_{ij}  = 
\langle\sigma_{ij}\rangle_{\cal D} 
\;\;;\;\;\Omega_{ij} = \langle\omega_{ij}\rangle_{\cal D}  \;\;.\eqno(B.5a,b,c)
$$
The averages $\langle\hat{\theta}\rangle_{\cal D} = 
\langle\nabla \cdot \vec u\rangle_{\cal D}$, 
$\langle\hat{\sigma}_{ij}\rangle_{\cal D}$ and 
$\langle\hat{\omega}_{ij}\rangle_{\cal D}$
then vanish by definition, and the expansion law (B.4) simplifies to 
$$
3{{\ddot a}_{\cal D} \over a_{\cal D}} + 4\pi G {M\over a_{\cal D}^3 } -
\Lambda = 
$$
$$
2(\Omega^2 - \Sigma^2 ) + 
\langle\nabla\cdot \lbrack \vec u (\nabla \cdot \vec u ) - 
(\vec u \cdot \nabla)\vec u \rbrack \rangle_{\cal D} \;\;.
\eqno(B.6)
$$
The last term in (B.6) is, via Gau{\ss}'s theorem, a surface integral 
over the boundary of $\cal D$. 
In case of a toroidal model we may choose $\cal D$ to be the whole torus.
Then, the background flow is {\it unique} (see Appendix A) and, since there
is no boundary, we obtain the global expansion law
(in agreement with the result of Brauer \etal loc. cit., eq.(2.15)):
$$
3{\ddot a \over a} + 4\pi G {M\over a^3 } -
\Lambda = 2\left(\Omega^2 - \Sigma^2 \right)\;\;.\eqno(B.6')
$$

\vfill\eject
\bye